\documentstyle[12pt,epsf]{article}
\begin{document}
\textwidth 13.6cm
\textheight 19.9cm
\noindent {\large \bf Effect of geometrical constraint on conformational
properties and adsorption transition of a semiflexible polymer chain} \\

\noindent {\it \large Pramod Kumar Mishra} \\
\noindent {\bf Department of Physics, DSB Campus, Kumaun University \\
Naini Tal-263 002, India} \\

\noindent {\bf Abstract} :
We analyze equilibrium properties and adsorption desorption phase transition behaviour of a linear semiflexible 
copolymer chain under constrained geometrical situation on square lattice in a good solvent. One dimensional
stair shaped line imposes geometrical constrain on the chain. Lattice model of fully directed self
avoiding walk is used to model the chain, semiflexibility of the chain is accounted by introducing
energy barrier for each bend of the chain. Exact expression of the partition function of the chain is
obtained using generating function technique for the cases, viz. (i) constrained copolymer
chain is in the bulk, (ii) constrained copolymer chain interacting with an impenetrable flat surface, (iii) constrained
copolymer chain interacting with constraint itself and (iv) general expression of the partition
function of the copolymer chain, interacting with a flat surface and geometrical constraint (stair shaped line).
We have compared bulk properties and adsorption desorption transition behaviour of a linear semiflexible homopolymer chain
without constraint to the case when the chain is constrained.

\vspace {.2cm}
\section{Introduction:}
Biomolecules ($DNA$ \& $proteins$) live in the crowded, constrained regime. Such molecules are soft object and therefore
can be easily squeezed into spaces that are much smaller than the natural size of the molecule in the bulk. For instance,
$actin$ filaments in $eukaryotic$ cell \cite{1} or $protein$ encapsulated in $Ecoli$ \cite{2} is found in nature are
examples of confined biomolecules serves the basis for understanding numerous phenomenon observed in polymer technology,
bio-technology and many other molecular processes occurring in the living cells. The conformational properties of single
biomolecules have attracted considerable attention in recent years due to developments in single molecule based observations
and experiments \cite{3,4,5,6,7,8}. Under confined geometrical condition, excluded volume effects and effect of geometrical constraint
compete with entropy of the molecule. Therefore, constraints can modify conformational properties and
adsorption desorption transition behaviours of macromolecules.

Behaviour of the flexible polymer molecule in a good solvent, confined to different geometries have been studied 
for past few years \cite{9,10}. For example, Brak and his coworkers \cite{11} used directed self avoiding
walk model to study behaviour of flexible polymer chain confined between two parallel walls on square lattice. However,
in present investigation we have considered a linear semiflexible copolymer chain on square lattice constrained by a one dimensional
stair shaped surface. For two dimensional space, surface is a line and copolymer chain is constrained by the such surface. The chain
is made of two type of monomers $A$ and $B$ distributed along the back bone of chain in a sequence $A-B-A-B\dots$. 
Therefore, it is a sequential or alternating copolymer chain.
We have considered sequential copolymer chain because it serve as a paradigmatic 
model of actually disordered macromolecule {\it e. g.} $proteins$.

To analyze the effect of constraint, we have used
fully directed self avoiding walk model introduced by Privmann {\it et. al} \cite{12} and used generating function technique 
to solve the model analytically.  
The results so obtained is used to discuss the behaviour of homopolymer chain in constrained geometry and also to compare results obtained
for conformational properties of the polymer chain for the case, when chain is in the bulk, in absence of one 
dimensional stair to the case when there is stair to constrain the chain. 
If constraint is an attractive surface, it contributes an energy $\epsilon_s$
($<0$) for each step of walk making on the surface. This leads
to an increased probability defined by a Boltzmann weight
$\omega=\exp(-\epsilon_s/k_BT)$ of making a step on the surface
($\epsilon_s < 0$ or $\omega > 1$, $T$ is temperature and
$k_B$ is the Boltzmann constant). 
The chain gets adsorbed on the surface at an appropriate  value of $\omega$ or $\epsilon_s$. 
Therefore, transition
between adsorbed and desorbed phases is marked by a critical value of adsorption
energy $\epsilon_s$ or $\omega_c$. 
The crossover
exponent $\phi$ at the adsorption transition point is defined as, $N_s \sim N^{\phi}$, where $N$ is the total
number of monomers in the chain while $N_s$ is the number of monomers adsorbed onto the surface. 
A comparison is also made to discuss effect of constraint on adsorption transition point.

The paper is organized as follows: In Sec. 2 lattice model of
directed self avoiding walk is described for a linear semiflexible sequential copolymer chain 
in good solvent. In sub-section 2.1, we have discussed behaviour 
of the copolymer chain in the bulk in presence of one dimensional stair shaped surface. The results
obtained in constrained geometry case is compared with those obtained in absence of constraint 
for semiflexible homopolymer chain.
Sub-section 2.2, is devoted to discuss adsorption of copolymer chain on a one dimensional flat surface 
in presence of constraint and results obtained is compared for the case when there is no constraint near the 
homopolymer and copolymer chains. In Sub-section 2.3, adsorption of copolymer chain on constraint is discussed. While
in sub-section 2.4, general expression of the partition function of copolymer chain which
is interacting with constraint and flat surface is discussed.
Finally, in Sec. 3 we summarize the results obtained.

\section{Model and method}
A lattice model of fully directed self-avoiding walk $\cite{12}$ is used to
model a semiflexible alternating copolymer chain in two dimensions and generating function technique has
been used to calculate partition function of the copolymer chain in presence of an impenetrable surface (line) having
shape like a stair (as shown schematically in figure 1). Since copolymer chain is fully directed 
therefore, walker is allowed to take steps along $+x$, $+y$ directions on the square lattice.
The stiffness of the chain is accounted by introducing an energy barrier for each bend in the walk of the chain. 
The stiffness weight $k=exp(-\beta\epsilon_{b})$ where $\beta=(k_BT)^{-1}$ is 
inverse of the temperature and $\epsilon_b(>0)$ is the energy associated with 
each bend of the walk of copolymer chain. For $k=1$ or $\epsilon_{b}=0$ the chain is said to be flexible and for 
$0<k<1$ or $0<\epsilon_{b} <\infty$ the chain is  
semiflexible. However, when $\epsilon_{b}\to\infty$ or $k\to0$, 
the copolymer chain is like a rigid rod.

Directed self avoiding walk model can be solved 
analytically and therefore gives exact value of conformational properties and 
adsorption transition point of the chain. However,  
directed walk model is restrictive in the sense that bending angle has same value {\it i. e.}
$90^0$ (for present study) for each bend and directedness of the walk introduces certain extent of stiffness in the 
polymer chain because all directions of the space are not treated equally by the walker. 

\begin{figure}[htbp] 
\centering 
\epsfxsize=14cm\epsfbox{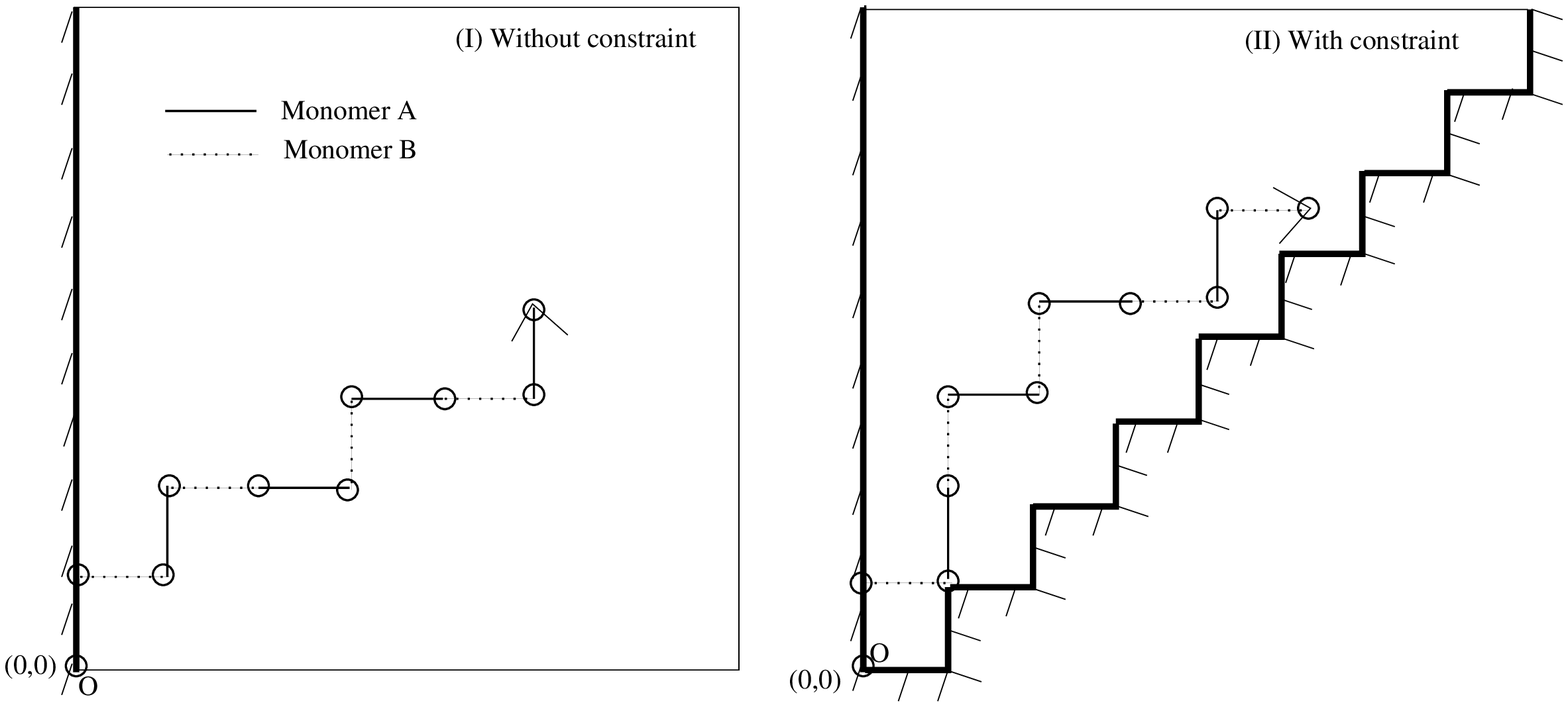} 
\caption{In this figure polymer chain is shown (I) in absence of constraint (one dimensional stair shaped surface) while in (II) there
is constraint. Walk of the chain starts from point O and directed along both the directions of space.}
\label{Figure1}
\end{figure} 

The partition function of a semiflexible sequential copolymer chain made of two type of monomers (A \& B) can be written as, 

\begin{equation}
Z(k,x_1,x_2)={\sum}^{N=\infty}_{N=0}\sum_{ all\hspace{0.07cm}walks\hspace{0.07cm}of\hspace{0.05cm}N\hspace{0.05cm}steps} {x_1}^{N/2}{x_2}^{N/2}k^{N_b}
\end{equation}
where, $N_b$ is the total number of bends in a self avoiding walk of $N$ steps (monomers),
$x_1$ and $x_2$ is the step fugacity of each of the A and B type monomers, respectively.
For the sake of mathematical simplicity we assume here and onwards $x_1=x_2=x$. Method
of analysis discussed in this paper can also be extended to the case when $x_1\ne x_2$. 

We consider an impenetrable line, located in the $x-y$ plane (as shown schematically in the figure 1) has shape like a stair, one end
of the copolymer chain is grafted to origin $O$. Since stair is impenetrable, therefore, it constrains the
polymer chain. We are interested in analyzing effect of this constrain on the conformation properties
of the copolymer chain in the bulk and also in calculating adsorption desorption phase transition point of the
copolymer chain in presence of constraint.

\subsection{\bf \it Copolymer chain in the bulk}

The partition function of a sequential semiflexible copolymer chain in the bulk is calculated using the
method of analysis discussed by Mishra {\it et al.}, $\cite{13}$ and 
components of the partition function $Z(k,x)$ of the copolymer chain in the bulk is written as follows:
\begin{equation}
X_0(k,x)= x+x^2k(1+Y+kX_0)
\end{equation}
where, $X_n(k,x)$ (here and onwards $n$=0,1,2,3,4...) is the sum of Boltzmann weight of all walks with 
first step along $+x$-direction and $n$ step distant from the stair
while $Y(k,x)$ is sum of Boltzmann weight of all the walks of the copolymer chain having first step along $+y$ direction.
Above equation can be re-written as,
\begin{equation}
X_0(k,x)= \frac{x(1+kx)}{1-x^2k^2}+\frac{x^2kY}{1-x^2k^2}
\end{equation}
Similarly, we have
\begin{equation}
\hspace{-2.5cm}X_n(k,x)=x+x^2+..+x^n+x^n\frac{x(1+kx)}{1-x^2k^2}+\large(kx+kx^2+...+kx^n+x^n\frac{x^2k}{1-x^2k^2}\large)Y
\end{equation}
and substituting $n=\infty$, we recover $X_\infty(k,x)=X(k,x)$ \cite{13},
\begin{equation}
X_\infty(k,x)= \frac{x}{1-x}+\frac{xkY'}{1-x}
\end{equation}
Where, $Y'$ is the other component of the partition function in absence of constraint. 
But, another component of the partition function of the semiflexible copolymer chain in presence of constraint is,
\begin{equation}
Y(k,x)=x+x(x+kX_1)+x^2(x+kX_2)+x^3(x+kX_3)+.......
\end{equation}
so that,
\begin{equation}
\hspace{-2.5cm}Y(k,x)=\frac{x(1-x^2)(1-x^2k^2)+kx^2(1-k^2x^2)+kx^2(1-x)(x+kx^2)}{(1-x)(1-x^2)(1-k^2x^2)-k^2x^2(1-x^2k^2)-k^2x^4(1-x)}
\end{equation}
Thus, the partition function of the copolymer chain 
in the bulk is written as,
\begin{equation}
Z(k,x)=X_0(k,x)+Y(k,x)
\end{equation}
such that
\begin{equation}
\hspace{-2.5cm}Z(k,x)=\frac{x(2+(-1+2k)x+(-2+k-2k^2)x^2+(1-2k+2k^2-2k^3)x^3)}{1-x-(1+2k^2)x^2+(1+k^2)x^3+k^4x^4}
\end{equation}
From singularity of the partition function $Z(k,x)$, we obtain critical value of step fugacity, $x_c$, needed for
polymerization of an infinitely long linear semiflexible copolymer chain. Since, we have considered $x_1=x_2=x$, we can 
treat this result for semiflexible homopolymer chain {\it i. e.} both the monomers of copolymer chain are identical. 
For, $k=1$, we have $x_c$(with constraint)=0.532089 while $x_c$(without constraint)=0.5 \cite{13} for flexible
homopolymer chain for fully directed self avoiding walk model on square lattice. Variation of $x_c$ with $k$ in
presence as well as absence of stair (constraint) is shown in figure 2.

Persistent length, $l_p$, of the chain can be calculated using definition of Mishra {\it et. al} \cite{13}, 
$l_p${\large $(=\frac{x\frac {\partial Log Z(k,x)}{\partial x}}{k\frac{\partial Log Z(k,x)}{\partial k}})$}, as average distance 
of the polymer chain between its two successive bends. The variation of the persistent length 
with $k$ is shown in figure 2. A comparison of $l_p$ is also made for the case when there no constraint \cite{13}.

\subsection{Adsorption on one dimensional flat line}
In order to analyze the effect of presence of constraint on adsorption transition point, we consider an impenetrable line
located at $x$=0 and calculated components of the partition function of the sequentional copolymer chain, which is interacting with
impenetrable flat line (while there is no attraction between monomer and constraint) 
using method of analysis used by Mishra {\it et. al}, we have component of the partition function along the surface as,
 
\begin{equation}
\hspace{-2.5cm}S^f(k,\omega_1,\omega_2,x)=s_1(1+s_2+kX_1)+s_1s_2(1+s_2+kX_2)+s_1^2s_2(1+s_2+kX_3)+\dots
\end{equation}
here $A$ type monomer of the copolymer chain is grafted at $O$, we have $s_1(=\omega_1x)$, $s_2=(\omega_2x)$ is the 
Boltzmann weight of $A$ and $B$ type monomers of the copolymer chain
with flat surface, respectively. While $X_n(k,x)$ is the component of the partition function
perpendicular to the flat surface {\it i. e.} along $x$-axis. The value of $X_n(k,x)$ we have used from Eq. (4) to calculate partition function,
$Z^f(k,\omega_1,\omega_2,x)$, of the copolymer chain (provided $s_1s_2<1$) as follows,
\begin{equation}
\hspace{-2.5cm}S^f(k,\omega_1,\omega_2,x)=\frac{s_1+s_1s_2}{1-s_1s_2}+\frac{s_1kx(1+s_2)}{(1-s_1s_2)(1-s_1s_2x^2)}+\frac{s_1s_2kx^2(1+s_1)}{(1-s_1s_2)(1-s_1s_2x^2)}+u_1+u_2
\end{equation}

\begin{figure}[htbp]
\epsfxsize=16cm
\centerline{\epsfbox{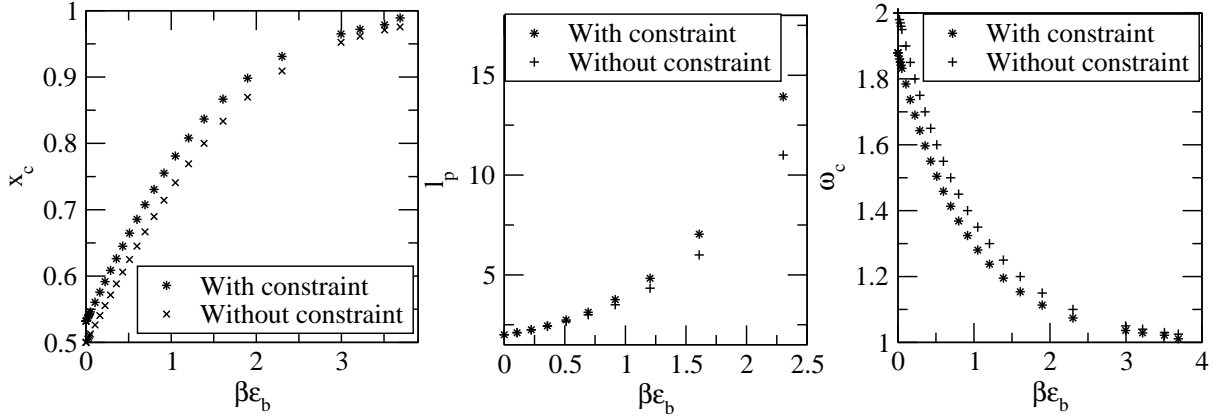}}
\caption{This figure compares the value of conformational properties of semiflexible
homopolymer chain when there is constraint (an impenetrable stair shaped surface) to that of the case if there 
is no constraint. We have also compared the value of adsorption desorption transition point of the semiflexible 
homopolymer chain on a flat surface in presence and absence of the constraint.}
\label{Figure2}
\end{figure}
where,

$u_1$={\Large$\frac{s_1kx(1+s_2x)(x+kx^2)}{(1-s_1s_2x^2)(1-x^2k^2)}$}

and

$u_2$={\Large$(\frac{s_1k^2x(1+s_2)}{(1-s_1s_2)(1-s_1s_2x^2)}+\frac{s_1s_2k^2x^2(1+s_1)}{(1-s_1s_2)(1-s_1s_2x^2)}+\frac{s_1(1+s_2x)k^2x^3}{(1-s_1s_2x^2)(1-k^2x^2)})$}Y
Therefore, partition function of the copolymer chain with $A$ type monomer grafted at $O$ can be calculated as,

\begin{equation}
Z^f(k,\omega_1,\omega_2,x)=S^f(k,\omega_1,\omega_2,x)+X_0(k,x)\hspace{2cm} (s_1s_2<1)
\end{equation}
\begin{equation}
\hspace{-2.5cm}Z^f(k,\omega_1,\omega_2,x)=\frac{f(k,\omega_1,\omega_2,x)}{(u_3(1-x-(1+2k^2)x^2+(1+k^2)x^3+k^4x^4)))}\hspace{1cm} (s_1s_2<1)
\end{equation}

where,

$u_3=(-1+s_1s_2)(-1+s_1s_2x^2)$

We are interested in the singularity of the partition function. Therefore, comparing the denominator of the partition
function $Z^f(k,\omega_1,\omega_2,x)$ equal to zero, we obtain critical value of monomer surface attraction
required for adsorption of copolymer chain, $\omega_{c1}=\frac{1}{\omega_{c2}x^2}$ on the flat surface. Effect of
constraint on adsorption transition of copolymer chain on flat surface is analyzed through figure (3). 
Replacing $s_1$ with $s_2$ in Eq. (13) we
could obtain partition function for copolymer having first monomer of $B$ type. The value of adsorption transition
point remain unchanged when grafted monomer of the copolymer chain is of $A$ or $B$ type. 
Substituting $\omega_{c1}=\omega_{c2}=\omega_c$, we get value of $\omega_c$ required for adsorption of homopolymer
semiflexible chain on an impenetrable line in presence of constraint. This result is shown in figure 2 and comparison
of this result with those obtained in absence of constraint \cite{13} is shown for different values of $k$.
\begin{figure}[htbp]
\epsfxsize=16cm
\centerline{\epsfbox{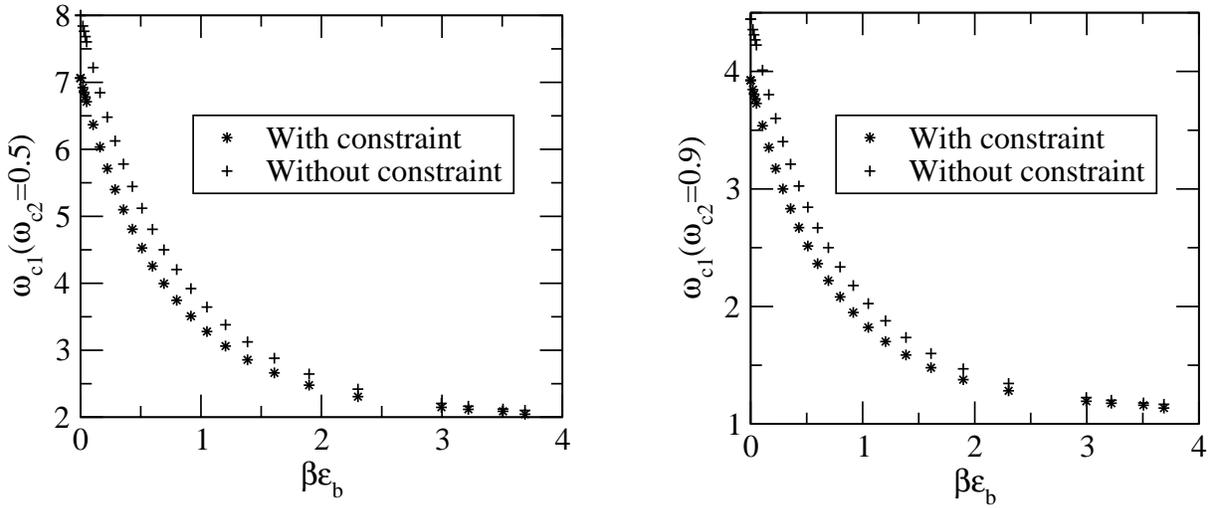}}
\caption{This figure compares the value of $\omega_{c1}$ required for adsorption of copolymer chain
on the flat surface for $\omega_{c2}=0.5,0.9$ in presence and absence of the constraint.}
\label{Figure3}
\end{figure}

\subsection{Adsorption of copolymer chain on constraint (one dimensional stair)}
The stair shaped one dimensional line (surface) is located in the $x-y$ plane as shown schematically in figure (1). The copolymer chain is interacting
with this surface. While, flat surface (line) located at $x=0$ is neutral for copolymer chain. The components of the partition function of
surface interacting copolymer chain can be calculated using method discussed above, as,

\begin{equation}
\hspace{-2.5cm}S_{0A}(k,\omega_3,\omega_4,x)= s_3+s_3s_4k(1+Y_{A}+s_3k)+s_3^2s_4k^2(1+Y_{A}+s_4k)+s_3^2s_4^2k^3(1+Y_{A}+s_3k)+\dots
\end{equation}

where, $s_3(=\omega_3x)$, $s_4(=\omega_4x)$ is the Boltzmann weight of interaction energy of $A$ and $B$ type monomers with stair shaped surface
respectively. $S_{nA}(k,\omega_3,\omega_4,x)$ [$A$ indicates that first monomer grafted at $O$ of the copolymer chain is of $A$ type] 
is the sum of Boltzmann weight of all walks having first step along $+x$-direction and $n$ step distant from the stair
while $Y_A(k,\omega_3,\omega_4,x)$ is sum of Boltzmann weight of all the walks of the surface interacting copolymer chain having first step 
along $+y$ direction. In the same fashion one can define $S_{nB}(k,\omega_3,\omega_4,x)$ and $Y_B(k,\omega_3,\omega_4,x)$. 
We assume that first monomer of copolymer chain is of $A$ type therefore for $s_3s_4<1$, we have,
\begin{equation}
S_{0A}(k,\omega_3,\omega_4,x)= \frac{s_3(1+s_4k)}{1-s_3s_4k^2}+\frac{s_3s_4kY_A}{1-s_3s_4k^2}
\end{equation}
also, for odd $n$=$p$,
\begin{equation}
\hspace{-2.5cm}S_{pA}(k,\omega_3,\omega_4,x)=x+x^2+..+x^n+x^n\frac{s_3(1+ks_4)}{1-s_3s_4k^2}+\large(kx+kx^2+...+kx^n+x^n\frac{s_3s_4k}{1-s_3s_4k^2}\large)Y_A
\end{equation}
and for even $n$=$q$,
\begin{equation}
\hspace{-2.5cm}S_{qA}(k,\omega_3,\omega_4,x)=x+x^2+..+x^n+x^n\frac{s_4(1+ks_3)}{1-s_3s_4k^2}+\large(kx+kx^2+...+kx^n+x^n\frac{s_3s_4k}{1-s_3s_4k^2}\large)Y_A
\end{equation}
The component of the partition function of surface interacting semiflexible copolymer chain along $y$-axis is,
\begin{equation}
Y_A(k,\omega_3,\omega_4,x)=x+x(x+kS_{1B})+x^2(x+kS_{2A})+x^3(x+kS_{3B})+\dots
\end{equation}
substituting the value of $S_{qA}(k,\omega_3,\omega_4,x)$ and $S_{pB}(k,\omega_3,\omega_4,x)$ in Eq. (18) we get,
\begin{equation}
\hspace{-2.5cm}Y_A(k,\omega_3,\omega_4,x)=\frac{x(1-x^2)(1-s_3s_4k^2)+kx^2(1-k^2s_3s_4)+kx^2(1-x)(s_3+ks_3s_4)}{(1-x)(1-x^2)(1-k^2s_3s_4)-k^2x^2(1-s_3s_4k^2)-k^2s_3s_4x^2(1-x)}
\end{equation}
Thus, we calculated partition function of copolymer chain interacting with a stair shaped impenetrable surface as,
\begin{equation}
Z^s(k,\omega_3,\omega_4,x)=S_{0A}(k,\omega_3,\omega_4,x)+Y_A(k,\omega_3,\omega_4,x)
\end{equation}
therefore, partition function of the copolymer chain which is interacting with an impenetrable stair shaped line is,
\begin{equation}
\hspace{-2.5cm}Z^s(k,\omega_3,\omega_4,x)=\frac{N^r(k,\omega_3,\omega_4,x)}{D^r(k,\omega_3,\omega_4,x)}
\end{equation}

where,

$N^r$=$x+kx^2-x^3-s_3(2k^3s_4x^2-(-1+x)^2(1+x)+k^2x(s_4+x-2s_4x)+k(-1+x)(s_4+s_4x+x^2))$

and,

$D^r=k^4s_3s_4x^2+(-1+x)^2(1+x)+k^2(s_3s_4(-1+x)-x^2)$

From singularity of the partition function $Z^s(k,\omega_3,\omega_4,x)$, we 
found that $\omega_3=\frac{-1+x+x^2+k^2x^2-x^3}{k^2\omega_4x^2(-1+x+k^2x^2)}$ 
({\it i. e.} on comparing $D^r(k,\omega_3,\omega_4,x)=0$) required for adsorption of semiflexible copolymer chain and it is independent of bending 
energy or stiffness $k$ of the polymer chain. 
This we have discussed through numerical results shown in figure (4) for $\omega_{c2}=0.5, 0.9 \& 1.0$. 
This is because, the polymer chain is directed towards the surface (constraint). 
Replacing $\omega_3$ and $\omega_4$ with $\omega_c$, we have adsorption 
transition point for homopolymer chain. The value of $\omega_c$ for homopolymer chain is close to unity (1$\pm$0.0001) 
for the values of $k$, $0.1\le k<1$.

\begin{figure}[htbp]
\epsfxsize=16cm
\centerline{\epsfbox{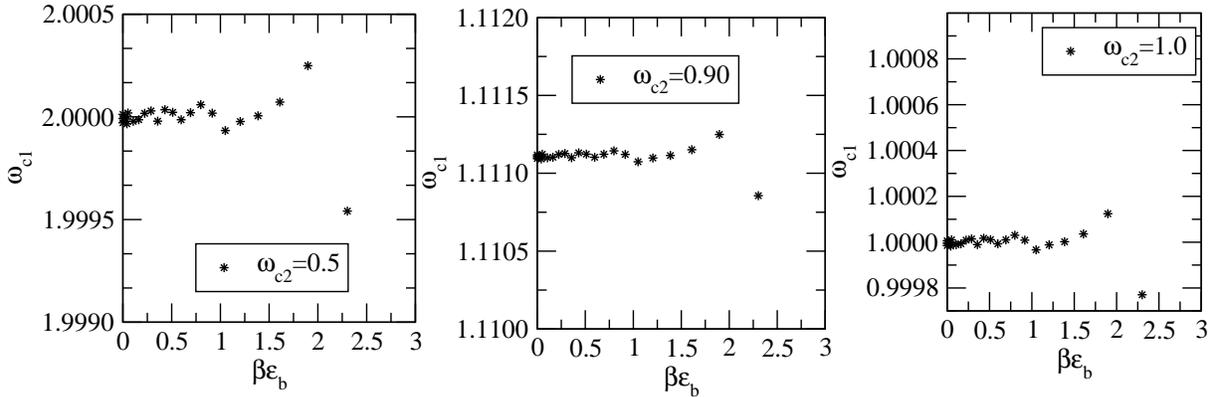}}
\caption{This figure is used to show that copolymer chain adsorption on constraint appears independent
of stiffness of the copolymer chain. However, in the very stiff chain limit adsorption of the chain on 
contsraint may not be possible.}
\label{Figure4}
\end{figure}

\subsection{General expression of the partition function}

We now derive expression of the partition function of the  sequential semiflexible linear 
copolymer chain for the case when copolymer chain is interacting with flat surface and constraint. 
The flat impenetrable surface (line) located at $x=0$ and another
surface is constraint (an impenetrable stair shaped line located along the line $x=y$; schematically shown in the figure (1)).
We assume that $A$ type monomer of copolymer chain is first monomer of the chain and one end of the chain is grafted at (0,0) or $O$. 
Let $S_A$ is the component of the partition function having first step on to the flat surface, then,
\begin{equation}
\hspace{-2.5cm}S_{A}(k,\omega_1,\omega_2,\omega_3,\omega_4,x)= s_1+s_1s_2k(1+S_{1B}+s_2k)+s_1^2s_2k^2(1+S_{2A}+s_2k)+\dots
\end{equation}

Substituting value of $S_{qA}$, $S_{pB}$ from Eqs. (16\& 17) and using following relation we have general expression of the partition function,
\begin{equation}
\hspace{-2.5cm}Z^{s\&f}(k,\omega_1,\omega_2,\omega_3,\omega_4,x)=S_{A}(k,\omega_1,\omega_2,\omega_3,\omega_4,x)+S_{0A}(k,\omega_3,\omega_4,x)
\end{equation}
\begin{equation}
\hspace{-2.5cm}Z^{s\&f}(k,\omega_1,\omega_2,\omega_3,\omega_4,x)=\frac{s_1(1+s_2)}{1-s_1s_2}+u_4+u_5-\frac{u_6}{u_7}+\frac{u_8}{u_{9}}\hspace{.5cm}(s_1s_2<1 ; s_3s_4<1)
\end{equation}

where,

\vspace{.2cm}

$u_4$={\Large$\frac{ks_3(1+ks_4)s_1x(1+s_2x)}{(-1+k^2s_3s_4)(-1+s_1s_2x^2)}$}

\vspace{.2cm}

$u_5$={\Large$\frac{ks_1(1+s_2)x(1+s_2x)}{(-1+s_1s_2)(-1+s_1s_2x^2)}$}

\vspace{.2cm}

$u_6=s_3(-(k^2(-1+s_4)x^2)+k^3s_4x^2-(-1+x)^2(1+x)+ks_4(-1+x^2))$

\vspace{.2cm}

$u_7=k^4s_3s_4x^2+(-1+x)^2(1+x)+k^2(s_3s_4(-1+x)-x^2)$

\vspace{.2cm}

$u_8=(k^2s_1x^2(-1-k^2s_3s_4(-1+x)+k^3s_3s_4x+k(-1+s_3(-1+x))x+x^2)(-1-s_2(1+x+s_1x)+s_3s_4((-1+s_1s_2)(1+s_2x)+k^2(1+s_2+s_2x+s_1s_2x))))$

\vspace{.2cm}

and

\vspace{.2cm}

$u_{9}=((-1+k^2s_3s_4)(-1+s_1s_2)(-1+s_1s_2x^4)(k^4s_3s_4x^2+(-1+x)^2(1+x)+k^2(s_3s_4(-1+x)-x^2)))$

\vspace{.2cm}

We call this relation (Eq. 24) as general expression of the partition function because from this expression we can recover 
expression of partition function discussed in above sub-sections by suitable substitutions. For example, replacing $s_1$, $s_2$,
$s_3$ and $s_4$ by x we have partition function of the copolymer chain in the bulk {\it i. e.} $Z(k,x)$. 

\section{Result and discussion}

We have considered a linear semiflexible sequential copolymer chain made of two type of monomers and fully directed
self avoiding walk model is used to model the copolymer chain on square lattice. 
Technique of generating function is used to calculate partition function of the chain in presence of constraint (an impenetrable one
dimensional line) analytically. 
The variation of
critical value of step fugacity required for polymerization of an infinitely long linear copolymer chain and 
persistent length under good solvent have been discussed. 
The results obtained with constraint is compared to case when there is no constraint. 
We have also compared the value of monomer surface attraction required for
adsorption of semiflexible copolymer chain on a flat surface and constraint. The surface is a one dimensional line
and walks of the chain is directed along both the directions of the space.

We have found that adsorption of the homopolymer chain on constraint occurs at a very small value of monomer and surface attraction. 
This is because, walks of the polymer
chain is directed along the stair shaped surface, walks of the chain follows curvature of the constraint and major 
contribution in the partition function of copolymer chain is due to walks lying on the constraint. 
In the case of stair shaped impenetrable surface, critical value of
monomer surface attraction required for the copolymer
chain adsorption appears independent of bending energy
of the chain (excluding the case of very stiff chains). While adsorption of copolymer and homopolymer chain on the flat surface 
is function of semiflexibility of the chain. The stiffer chain adsorption occurs at a smaller
value of monomer surface attraction than the flexible polymer chain.
These features of adsorption
on an impenetrable flat and curved surfaces are similar to the semiflexible polymer adsorption $\cite{13,14,15,16,17}$.

\end{document}